\documentstyle[12pt,here,epsf,a41,rotate,dcolumn]{article}

\newcommand{\gsim}{\raisebox{-0.07cm}{$\, \stackrel{>}{{\scriptstyle
\sim}}\, $}}

\begin{document}
\newcolumntype{-}{D{-}{\mbox{--}}{-1}}
\begin{center}
{\LARGE\bf The polarised gluon density  from di-jet events in DIS at
a polarised HERA}

\vspace{1cm}
{G. R\"adel$^a$, A. De Roeck$^{b}$, M. Maul$^c$}

\vspace*{1cm}
{\it $^a$CERN - Div.\ PPE, 1211 Gen\`eve 23, Switzerland.}\\
\vspace*{3mm}
{\it $^b$DESY - FH1, Notkestr.\ 85, 22603 Hamburg, Germany}\\
\vspace*{3mm}
{\it $^c$Institut f\"ur Theoretische Physik der Universit\"at Regensburg,
93040 Regensburg, Germany}\\

\vspace*{2cm}

\end{center}

\begin{abstract}
We present a possible direct measurement of the polarised gluon density
$\Delta G(x)$ in LO from di-jet production in polarised deep inelastic $ep$
scattering, assuming the kinematics of the HERA collider. 
We show the sensitivity to the $x$-dependence of $\Delta G(x)$ 
and to the first moment $\int \Delta G(x) dx$ in the range $0.002 < x < 0.2$,
assuming the electron and proton beam of HERA being polarised to 70\% and
an integrated luminosity of at least 200~pb$^{-1}$. We include in our
study hadronisation and higher order effects, as well as realistic
detector smearing and acceptance. We find that the statistical and systematic
uncertainties are small enough to distinguish between different parametrizations for $\Delta G(x)$, which all are in accordance with present data.
We stress that at HERA an $x$-range could be measured, that is not accessible
to any other present or proposed experiment.
\end{abstract}

\section{Introduction}

\vspace{1mm}
\noindent
The precise study of the nucleon spin structure has evolved over the last 
years into a broad field allowing to test many aspects of QCD.
The surprising EMC~\cite{EMC} result, 
that the quarks carry only a small fraction of the 
nucleon spin, has been confirmed  by new high precision measurements.
This dramatic improvement in quality of the data and  theoretical analysis
has lead to a generally accepted range of polarised 
parton distribution parametrizations which imply  that the quarks 
carry only about 30\% of the nucleon spin. Nearly all models predict 
a substantial polarisation for  both the gluons and the strange quarks which 
has to be confirmed by direct experimental tests before the present 
standard interpretation of the data can be regarded as established.
The polarised gluon distribution is of special interest since it could be 
surprisingly large.

In the next to leading order (NLO) evolution equations the quark and gluon 
distributions mix, hence polarised gluons also contribute to $g_1(x,Q^2)$.
The quality of the present $g_1$ data allow for QCD fits to be made, 
and $\Delta G$ to be extracted. The precision is however rather poor, 
and only some information on the first moment on $\Delta G$ is obtained.
Therefore  the hunting of $\Delta G$ by direct measurements is one of the 
key issues in polarised scattering physics for the next foreseeable future.

The unpolarised gluon distribution has been studied at the $ep$
collider HERA. The large centre of mass energy ($ \sqrt{s} = 300$) GeV, 
resulting from 27.5 GeV electrons colliding with 820 GeV protons, allows for 
several  techniques to be used. 
So far the gluon distribution at HERA has been accessed via
scaling violations of $F_2$, di-jet production, charm production and 
exclusive vector meson production.
While the first method gives an indirect measurement of the gluon, like
for the NLO analysis of $g_1$, for the other methods the gluon enters 
directly at
the Born level. In this paper we will
use 
the method of extracting the gluon via  di-jet events rates.

\begin{figure}[h]
\hfil \epsffile{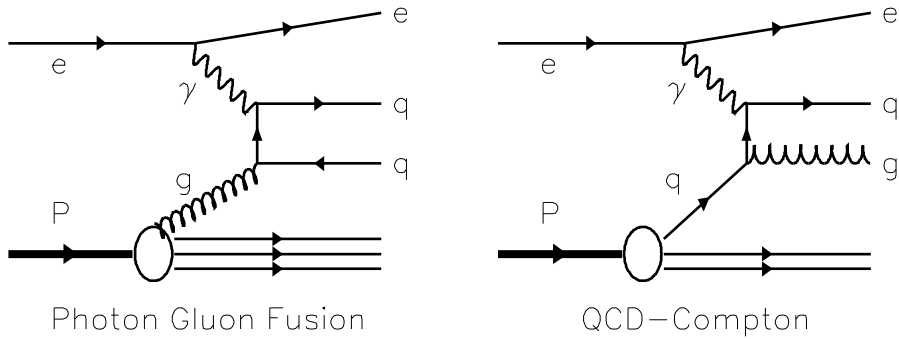} \hfil
\vspace{-1cm}
 \hspace*{12cm}
\caption{Feynman diagrams for the di-jet cross sections at LO,
the Photon-Gluon Fusion (PGF) process (left) and the QCD-Compton process
(right). }
\label{fig-feyn}
\end{figure}

In LO two diagrams can lead to di-jet events, shown in Fig.~1.
These are the Photon-Gluon Fusion process (PGF) and the QCD-Compton
process (QCDC). The PGF process is directly sensitive to the gluon density,
while the QCDC process is sensitive to the quark densities and constitutes the 
background. The H1 collaboration has performed an analysis of di-jets
to extract the PGF contribution and thus  the gluon 
distribution~\cite{H1}. Presently both the H1 an ZEUS collaborations 
attempt to extract the gluon distribution at NLO from
di-jets event rates~\cite{H1NLO,ZEUSNLO}.

If both beams at HERA would be  polarised, this method could be used
to extract $\Delta G$.
Due to the Sokolov-Ternov effect the electron beam 
gets transversely polarised in the 
machine.
Spin rotators 
can flip 
transverse into longitudinal polarisation,
which is more useful for physics studies.
The only 
possibility for 
a polarised proton beam at HERA 
is to start from a polarised source
and accelerate and store the beam, keeping the 
polarisation on the way~\cite{Barber}.  First 
feasibility studies indicate the possibility
of such a scenario in case  the accelerators get
upgraded with partial and full Siberian snakes. For this report it
is assumed that a  polarisation
of 70\% can be reached for both beams, and that 
the luminosity will be as large as for the
unpolarised case (roughly 200 to 500 pb$^{-1}$, integrated over several years)

First studies on extracting $\Delta G $ from 
di-jet event rates were made in~\cite{Feltesse,ourpaper}.
In this paper
 we make a full Monte Carlo simulation of the signal and background
processes, include hadronisation, higher order effects via parton
showers, and detector effects. Starting from three different sets of 
polarised gluon distributions, shown in Fig.~2, 
we check the sensitivity of the 
measurements and extract $\Delta G(x)$. 
These distributions are the Gehrmann-Stirling 
(GS) sets A and C~\cite{GS}, which
result from a QCD analysis of $g_1$ data, and the instanton-gluon
distribution~\cite{Kochelev}. The latter results from a calculation
of the polarised parton distribution in the Instanton Liquid 
Model~\cite{Instanton}.
The distributions shown in Fig.~2, purposely selected,
indicate how poorly $\Delta G(x)$ is constrained by 
 the present polarised data. All of these distributions are
compatible with the  
available data, stressing the need for direct measurements
of $\Delta G(x)$. The GS-A and GS-C distribution show a similar small
$x$ behaviour, but differ considerably in the region around $x \sim 0.1$.
The GS-C distribution is negative for this $x$ region.
The instanton-gluon is quite different from the GS sets. It remains 
negative over the full $x$ range. The latter gluon is used in combination with 
the GS-A quark distributions for the study in this paper.
For the unpolarised parton density functions the parametrizations 
of Gl\"uck, Reya and Vogt in LO were used~\cite{GRV94}.

\begin{figure}[t]
\epsfxsize=9cm
\hfil \epsffile{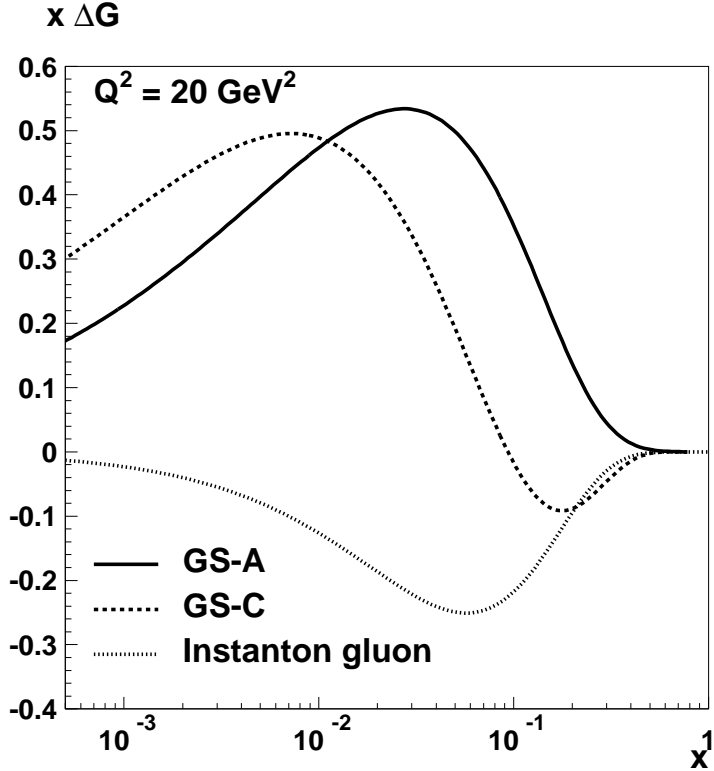} \hfil

\vspace{-2cm}
 \hspace*{12cm}
\caption{Different parametrizations of the polarised gluon density
as a function of $x$ used in this analysis, described in the text.}
\label{input}
\end{figure}

\section {Jet cross sections and 
Monte Carlo programs}

\vspace{1mm}
\noindent
Deep inelastic electron-proton scattering with several partons
in the final state,
\begin{equation}
e^-(l) + p(P) \rightarrow  e^-(l^\prime)+
\mbox{remnant}(p_r) +
\mbox{parton} \,\,1 (p_1) +
\ldots
+\mbox{parton}\,\, n (p_n)
\label{eq1}
\end{equation}
proceeds via the exchange of an
intermediate vector boson $V=\gamma^*, Z$. $Z$-exchange and $\gamma^*/Z$
interference become only important at large $Q^2$ ($> 1000 $ GeV$^2$)
and are neglected 
in the following.
 We denote the momentum of the incoming proton by $P$, the
momentum of the virtual photon, $\gamma^{\ast}$, by $q=l-l^\prime$, 
(minus) its absolute square by $Q^2$,
and use the standard scaling variables Bjorken-$x$
$x={Q^2}/({2P\cdot q})$ and inelasticity $y={P\cdot q}/{P\cdot l}$.
The general structure of the unpolarised
 $n$-jet cross section in DIS is given by
\begin{equation}
d\sigma^{had}[n\mbox{-jet}] = \sum_a
\int dx_a \,\,f_a(x_a,\mu_F^2)\,\,\, d\hat{\sigma}^a(p=x_a P, 
\alpha_s(\mu_R^2), \mu_R^2, \mu_F^2)
\label{eq2}
\end{equation}
where the sum runs over incident partons $a=q,\bar{q},g$ which carry 
a fraction $x_a$ of the proton momentum.
$\hat{\sigma}^a$ denotes the partonic cross section from which collinear 
initial state singularities have been factorized out 
(in  next-to-leading order (NLO)) at a scale $\mu_F$ and 
implicitly included in the scale dependent parton densities 
$f_a(x_a,\mu_F^2)$. 
For longitudinally polarised lepton-hadron scattering, the hadronic ($n$-jet) 
cross section is obtained from Eq.~(\ref{eq2}) by replacing
$(\sigma^{had}, f_a\,\, , \hat{\sigma}^a)\rightarrow
 (\Delta\sigma^{had}, \Delta f_a, \Delta\hat{\sigma}^a)$.
The polarised hadronic cross section is defined by
$\Delta\sigma^{had}\equiv\sigma^{had}_{\uparrow\downarrow}
                   -\sigma^{had}_{\uparrow\uparrow}$, where the
left arrow in the subscript denotes the polarisation of the 
incoming lepton with respect to the direction of its momentum.
The right arrow stands for the polarisation of the proton parallel 
or anti-parallel to the polarisation of the incoming lepton.
The polarised parton distributions are defined by
$\Delta f_a(x_a,\mu_F^2)
\equiv f_{a \uparrow}(x_a,\mu_F^2)-f_{a \downarrow}(x_a,\mu_F^2)$.
Here, $f_{a \uparrow} (f_{a \downarrow})$ denotes 
the probability to find a parton $a$ 
in the longitudinally polarised  
proton whose spin is aligned (anti-aligned) to the proton's spin.
$\Delta\hat{\sigma}^a$ is the corresponding polarised
partonic cross section.
The subprocesses  
$\gamma^\ast +q \rightarrow q + g$, 
$\gamma^\ast +\bar q \rightarrow \bar q + g$,  
$\gamma^\ast+g \rightarrow q + \bar{q}$  
contribute to the di-jet cross section (Fig.~\ref{fig-feyn}).
The photon-gluon fusion subprocess
$\gamma^\ast+g \rightarrow q + \bar{q}$  dominates the
di-jet cross section at low Bjorken-$x$ 
for unpolarised protons 
(see below) and allows for a  direct
measurement of the gluon density in the proton.
The full NLO corrections for di-jet production in unpolarised
lepton-hadron scattering  are 
implemented in the 
$ep \rightarrow n \rm  -jets$ event generator MEPJET~\cite{Mirkes} 
which allows to analyse 
arbitrary jet definition schemes and 
general cuts in terms of parton 4-momenta.
Recently, MEPJET has been extended to NLO for polarised 
scattering~\cite{2jets-NLO}, and the NLO QCD corrections are found to be 
moderate.\\
In LO the total unpolarised di-jet cross section is the sum 
of the contributions from photon gluon fusion processes, $\sigma_{di\mbox{-}jet}^{PGF},$ and QCD-Compton
scattering, $\sigma_{di\mbox{-}jet}^{QCDC}$ and can be written as:
\begin{equation}
\sigma_{di\mbox{-}jet} =\sigma_{di\mbox{-}jet}^{PGF} +\sigma_{di\mbox{-}jet}^{QCDC} = A \,\, G + B\,\,  q
\end{equation}
where $G$ and $q$ are the gluon and quark densities and $A$ and $B$ can
be calculated in perturbative QCD. Similarly, for the polarised case we
can write:
\begin{equation}
\Delta \sigma_{di\mbox{-}jet} = \sigma_{di\mbox{-}jet}^{\uparrow\downarrow} - \sigma_{di\mbox{-}jet}^{\uparrow\uparrow} = a\,\,  \Delta G + b\,\, \Delta q 
\end{equation}
with $ \Delta G$ and $\Delta q$ being the polarised gluon and quark
densities.
The di-jet asymmetry is therefore sensitive to $\Delta G/G$, especially
at low $x$ where the PGF cross section dominates.
\begin{equation}
A_{di\mbox{-}jet} = \frac{\Delta \sigma_{di\mbox{-}jet}}{2\, \sigma_{di\mbox{-}jet}} = 
{\cal A}\,\,\frac{\Delta G}{G}\,\,\, \frac{\sigma_{di\mbox{-}jet}^{PGF}}{2\, \sigma_{di\mbox{-}jet}}
+ {\cal B}\,\,\frac{\Delta q}{q}\,\,\, \frac{1}{2}\, (1-\frac{\sigma_{di\mbox{-}jet}^{PGF}}{\sigma_{di\mbox{-}jet}}),
\label{eq-A}
\end{equation}
with ${\cal A} \equiv a/A$ and ${\cal B} \equiv b/B$. 
The experimentally accessible asymmetry $A_{meas}$ is smaller
than  $A_{di\mbox{-}jet}$ due to the  incomplete polarisations 
of the electron and proton beams, given by $P_e, P_p$, and the depolarisation
of the $\gamma^*$ with respect to the electron. The latter effect
is described by the depolarisation factor $D=(y(2-y))/(y^2+2(1-y)(1+R))$,
where $y$ is the inelasticity and $R$ is the ratio of longitudinal and
transverse $\gamma^* p$ cross sections.
\begin{equation}
A_{meas} =  \frac{N^{\uparrow \downarrow}-N^{\uparrow \uparrow}}
{N^{\uparrow \downarrow}+N^{\uparrow \uparrow}} = P_e P_p D A_{di\mbox{-}jet}
\end{equation}
The quantities $N^{\uparrow \downarrow}$ ($N^{\uparrow \uparrow}$)
are the total number of observed di-jet events 
($N^{\uparrow \downarrow}=N^{\uparrow \downarrow}_{PGF}
+N^{\uparrow \downarrow}_{QCDC}$) for the case
that proton and electron spin are antiparallel (parallel), respectively. 
The kinematic quantities to describe the PGF process are the momentum
fraction of the proton carried by the gluon $x_g$, the four-momentum transfer
$Q^2$ and the square of the invariant
mass of the two jets $s_{ij}$. They are related to the Bjorken-$x$,
$x$, by:
$$x_g=x (1+\frac{s_{ij}}{Q^2}).$$
For $s_{ij} > 100~{\rm GeV}^2$, and 
in the $Q^2$ range relevant for HERA of $5<Q^2<100$~GeV$^{2}$,
$x_g$ is larger than Bjorken-$x$ by about an order of magnitude
and the accessible range at
HERA is therefore 
about $0.002 < x_g < 0.2$. H1 has demonstrated that in this region 
the unpolarised gluon density $G(x)$ can be extracted from di-jet cross sections~\cite{H1}.\\

The program MEPJET has been used to study  di-jet
production in (un)polarised DIS at (N)LO at the level of 
parton-jets~\cite{Feltesse,ourpaper,Mirkes}. 
To perform a study for a possible future measurement it is, however, desirable to include also hadronisation and detector effects. 
Therefore in this study we use the Monte Carlo event generator program 
PEPSI~6.5~\cite{PEPSI1,PEPSI65}.
 It is a full, LO lepton-nucleon scattering 
Monte Carlo program based on LEPTO 6.5~\cite{LEPTO65}
 for unpolarised and polarised interactions, including
fragmentation, and unpolarised parton showers to simulate higher order effects.
\vspace{-1cm}
\begin{figure}[H]
\epsfxsize=15cm
\hfil \epsffile{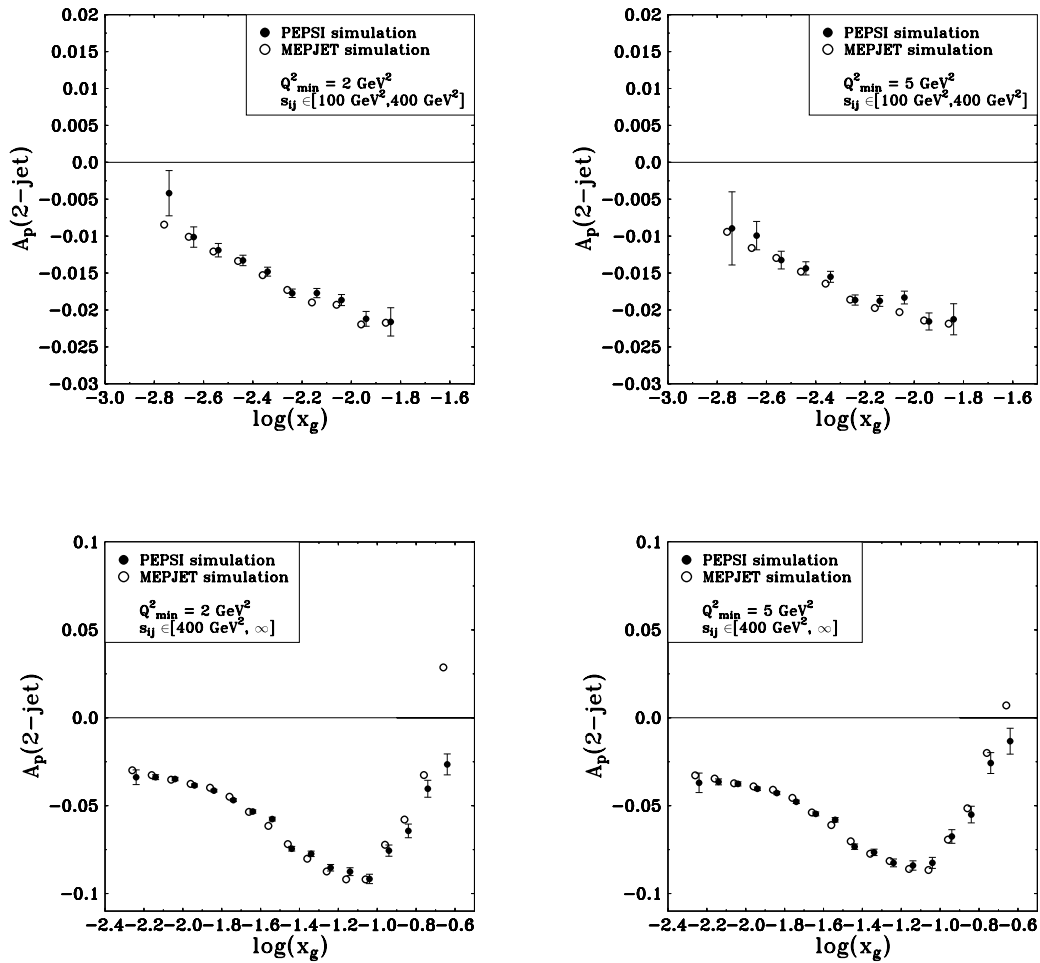} \hfil
\vspace{-1cm}
 \hspace*{12cm}
\caption{Comparison of the di-jet cross sections obtained with the
MC programs MEPJET and PEPSI~6.5. The cross sections are shown for different 
bins of $s_{ij}$ and for a lower $Q^2$-cut of 2 and 5 GeV$^2$.
The error bars for the PEPSI study represent the statistical uncertainty
corresponding to the generated number of events.}
\label{fig-PEPJET}
\end{figure}
A comparison of PEPSI and MEPJET, used with identical conditions 
and cuts, shows that,
at parton level,
both programs give very similar results for di-jet cross sections
and asymmetries for the
HERA kinematic range~\cite{PEPSI-MEPJET}. 
As an example Fig.~\ref{fig-PEPJET} shows the comparison for
the di-jet asymmetry ($A_p \mbox{(2-jets)} \equiv D\, A_{di\mbox{-}jet}$)
as function of $x_g$. The agreement between the MEPJET and PEPSI calculation
is very good.
The calculations were done for HERA energies, 820 GeV protons and 27.5 GeV
electrons. The kinematic range was restricted to $0.3 < y < 0.8$
and $Q^2 < 100~{\rm GeV}^2$.  The minimum $Q^2$ was varied from 2~GeV$^2$ 
(plots on the left) 
to 5~GeV$^2$ (plots on the right).
In PEPSI the so-called   $z\mbox{-}\hat{s}$ recombination scheme~\cite{jetschemes}
has been used to define the phase space available for the LO matrix elements.
The parameters for this scheme, $z_{min}=0.04$ and $\hat{s}_{min}=100~{\rm GeV}^2$, were chosen such that the phase space region for di-jet events using
a cone jet scheme for di-jets 
with $p_t > 5$~GeV and $s_{ij} > 100~{\rm GeV}^2$ was not
affected. 
$z_{min}$ is the minimum of $z=(P\cdot p_{jet})/(P\cdot q)$ for the two
jet momenta $p_{jet}$ in a di-jet event. The variable 
$\hat{s}$ is defined via $\hat{s} \equiv (p+q)^2$, where $p$ is the momentum of the incoming quark. 
For the jet detection a cone jet algorithm was used with $R_{min}=1$,
$R_{min}$ being
the minimal distance which two partons must have in order to belong to different jets. $R$ is given by: $R=\sqrt{(\Delta \eta)^2 + (\Delta \phi)^2}$ with
$ \eta$ being the pseudo rapidity  and $ \phi$ being the azimuthal angle in
the laboratory frame.
The two upper plots show the  asymmetries for events with
$100 < s_{ij} < 400~{\rm GeV}^2$, for the two lower plots $s_{ij}>400~{\rm GeV}^2$. The division into two $s_{ij}$ bins was made, because studies for
the unpolarised di-jet cross sections~\cite{Mirkes,Feltesse}
 have shown that NLO
corrections are expected to be small above $s_{ij}\gsim 400~{\rm GeV}^2$. 

In ref.~\cite{PEPSI-MEPJET} a study  was made to further optimize the cuts 
in order to get a better sensitivity to $\Delta G$. They concluded that
the cuts used in this analysis are already very close to the 
optimum choice.

Recently, some disagreement at low $Q^2$
 has been reported between the newly, more 
precise, measured jet cross sections and 
the NLO calculations at HERA, using the 
cone jet algorithm~\cite{rosen}. 
This  discrepancy may hint towards a 'resolved' 
photon component in the data, and is presently under study. We expect
however
that at the 
time the measurement described in this paper can be made, this matter 
will be settled and will have the effect of an additional small 
background to be subtracted from the di-jet event rates, in order to 
access the gluon distribution.

\section {Measured asymmetries}

\vspace{1mm}
\noindent
We  present a detailed study using the PEPSI program  on
the expected size of the measurable asymmetries for
di-jet production at HERA. We  show the influence of parton showers,
which 
simulate higher order effects, and hadronisation and detector effects. We also
show the sensitivity of the measurement to different polarised gluon
distributions.

\begin{figure}[h]
\epsfxsize=12cm
\hfil {\epsffile{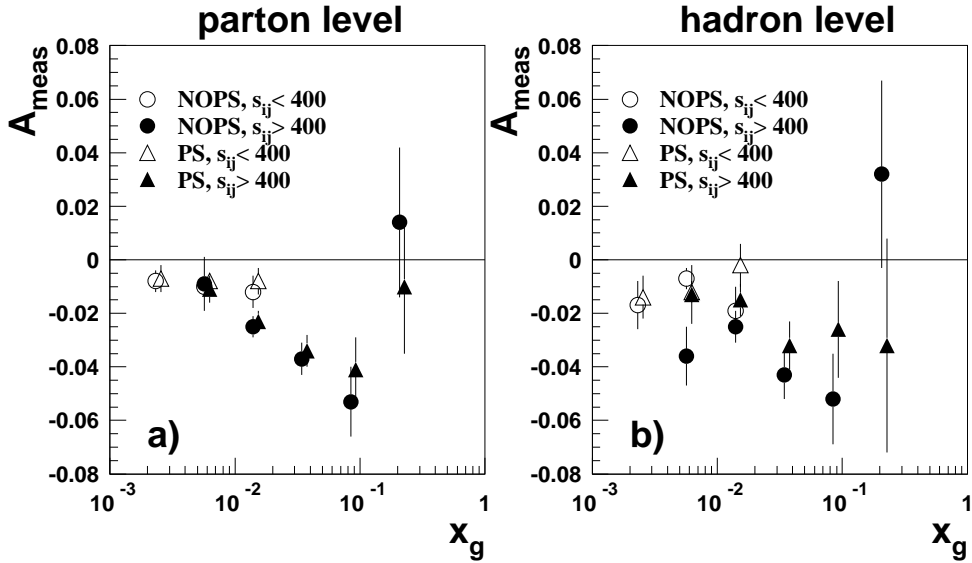}} \hfil
\vspace{-5cm}
 \hspace*{8cm}
\caption{ Expected measured asymmetries for di-jet events as a function
of $x_g$ calculated with PEPSI on the parton level (a) and detector level
(b). For each case the asymmetries are shown with (PS) and without
(NOPS) parton showers
for two different ranges of $s_{ij}$. The assumed integrated luminosity is
200~pb$^{-1}$. The input polarised gluon distribution is GS-A.}  
\label{fig-asy-ph}
\end{figure}

The kinematic cuts applied for this study
are similar to the ones discussed in the previous section,
$5 < Q^2 < 100$~GeV$^2$ and $0.3 < y < 0.85$. Again two
bins of $s_{ij}$ were analysed with $100 < s_{ij} < 400$~GeV$^2$ and
$s_{ij}>400$~GeV$^2$, respectively. Jets are defined using
 the cone scheme, are
required to have a $p_t>5$~GeV and are restricted to the acceptance 
of a typical existing HERA detector by $|\eta_{jet}| < 2.8$, were
$\eta_{jet}$ is the pseudo-rapidity in the laboratory system. 
The expected measurable asymmetry for the input polarised gluon density
GS-A, assuming the beam polarisations $P_e=P_p=0.7$
and 200~pb$^{-1}$ for the luminosity,
is shown in Fig.~\ref{fig-asy-ph}a
at the parton level. The expected asymmetry
is negative and 
of the order of a few $\%$. Jets induced by parton showers
 tend to reduce
the size of the asymmetry. 
This is due to the  fact that parton showers can 
produce a hard jet, which is then misidentified as a PGF induced one.
This, rather small 
reduction on parton level,  is more pronounced, if
 hadronisation and detector smearing effects are included (see 
Fig.~\ref{fig-asy-ph}b).
The reason for this is that both effects broaden the jets and therefore
the measured $p_t$ which is related to the energy in the cone of
fixed size is smaller than the $p_t$ of the parton jet.
The reconstruction of the kinematics of the event ($s_{ij}$, $x_g$)
is influenced and the correlation with the parton jets is reduced.
For the hadronisation the Lund fragmentation model, implemented in 
JETSET~\cite{JETSET},
was used and an energy resolution for the hadronic  calorimeter
of 
$\Delta E_{had}/E_{had} =
 0.5 / \sqrt{E_{had}[{\rm GeV}]}$ was assumed. 
\vspace{-3cm}
The results were cross checked using a realistic simulation program of
the H1 calorimeter~\cite{guillermo}, which takes into account
the energy resolution, 
the absolute energy scale and  dead material
in the detector. 
In order to optimize the signal to background ratio
cuts
were introduced demanding the two jets to be produced with a restricted
 difference in 
pseudo-rapidity  and back to back in azimuth,
as it is expected for real PGF events:
$|\eta_{jet 1}-\eta_{jet 2}| < 2$ and $ 150^{\circ} < \phi_{jet 1} - \phi_{jet 2} < 210^{\circ}$. 
After all these cuts for 100~pb$^{-1}$ about 70,000 di-jet
 events are selected. The ratio of QCDC - PGF events is in the order of
1:6. 
The average $Q^2$ of this event sample 
 is very close to 20~GeV$^{2}$
therefore  results for  $\Delta G$ are presented at this value. \\

All  cuts are applied in the asymmetry shown in Fig.~\ref{fig-asy-ph}b.
Although the asymmetries are smaller due to the parton showers, and the
statistics is reduced compared to the result on parton level, the
expected asymmetry is still large enough to allow a statistically significant
measurement for 200~pb$^{-1}$. 
These asymmetries form the basis of the studies in this paper.
Due to the split-up in the $s_{ij}$-bins, for the second and third $x_g$
bin there are two measurements.
For simplicity we choose in the following one measurement per $x_g$-bin,
i.e.\ the one
with the better significance. However, in principal the other points
could be used as well, and add to the statistical significance.
Table~\ref{tab-asys} shows the expected asymmetries $A_{meas}$ and
their statistical errors $\delta (A)$ for
the six $x_g$-bins shown in Fig.~\ref{fig-asy-ph}. The two lowest
$x_g$-bins correspond to $100 < s_{ij} < 400$~GeV$^2$. 
Also shown is $A_{corr}$ which is defined as $A_{corr} = \frac{N^{\uparrow \downarrow}_{PGF}-N^{\uparrow \uparrow}_{PGF}}
{N^{\uparrow \downarrow}+N^{\uparrow \uparrow}}$ and corresponds to 
the first term in the sum of Eq.~\ref{eq-A},
which is the part sensitive to $\Delta G/G.$ In other words, it is the 
measured asymmetry corrected for the QCDC contribution.
The numbers in Table~\ref{tab-asys} show that a significant contribution
from QCDC processes is expected only for the two highest $x_g$-bins.

\begin{table}
\hfil
\hspace{-4.5cm}
\begin{tabular}{||c|c|c|c||}
\hline\hline
 & \multicolumn{3}{|c||}{$ 5 < Q^2 < 100~{\rm GeV}^2$} \\
$x_g$ & $A_{meas}$ & $A_{corr}$ & $\delta(A)$ \\
\hline \hline
0.002 & -0.016 & -0.016 & 0.008 \\
0.006 & -0.012 & -0.012 & 0.004  \\
0.014 & -0.015 & -0.018 & 0.005 \\
0.034 & -0.032 & -0.032 & 0.009\\
0.084 & -0.026 & -0.047 & 0.018 \\
0.207 & -0.032 & -0.069 & 0.040 \\
\hline\hline
\end{tabular}
\hfil
\vspace{-4.75cm}
\begin{flushright}
\hspace{-5cm}
\begin{tabular}{||c|c|c||c|c||}
\hline\hline
 & \multicolumn{2}{|c||}{$ 2 < Q^2 < 10~{\rm GeV}^2$} &  \multicolumn{2}{|c||}{$ 10 < Q^
2 < 100~{\rm GeV}^2$}\\
$x_g$ & $A_{corr}$ & $\delta(A)$  & $A_{corr}$ & $\delta(A)$ \\
\hline \hline
0.002 & -0.009 & 0.009 & -0.006 & 0.010\\
0.006 & -0.010 & 0.005 & -0.014 & 0.005\\
0.014 & -0.017 & 0.007 & -0.026 & 0.007\\
0.034 & -0.042 & 0.010 & -0.037 & 0.010\\
0.084 & -0.045 & 0.021 & -0.073 & 0.021\\
0.207 & -0.096 & 0.046 & -0.008 & 0.050\\
\hline\hline
\end{tabular}
\end{flushright}
\hfil
\caption{Expected measured asymmetries from di-jet events at HERA and
 background corrected asymmetries with
statistical errors corresponding to an integrated luminosity of
200~pb$^{-1}$. The right table shows the background corrected asymmetries for
two different $Q^2$ ranges.}
\label{tab-asys}
\end{table}

The right part of Table~\ref{tab-asys} shows results for $A_{corr}$
and its statistical error if the low $Q^2$ cut is released to 2~GeV$^2$
and the data are divided into to $Q^2$ bins. The mean $Q^2$ values for the
two bins are 4.5~GeV$^2$ and 30~GeV$^2$, respectively.

\begin{figure}[t]
\epsfxsize=9cm
\hfil \epsffile{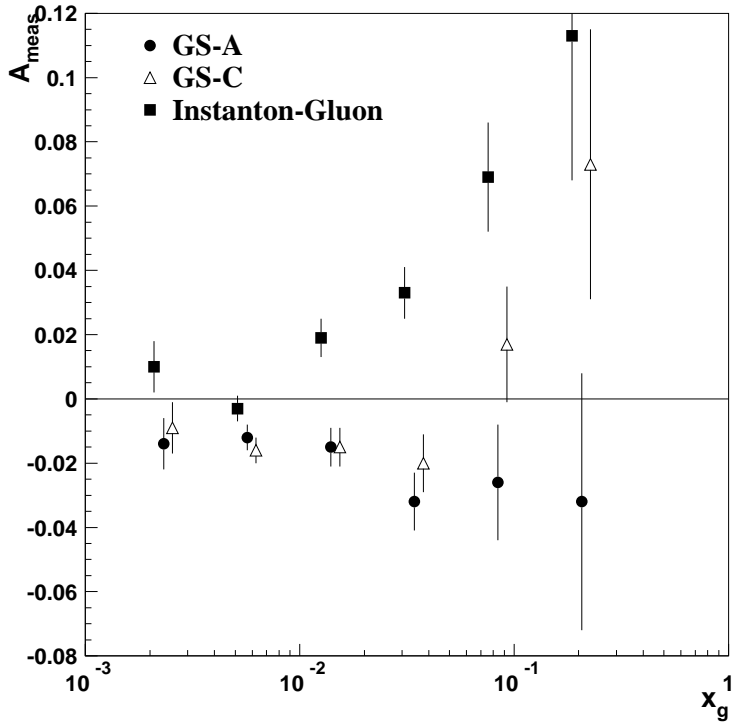} \hfil
\vspace{-2cm}
 \hspace*{12cm}
\caption{Expected measured asymmetries for di-jet events for different 
input polarised gluon densities. The assumed integrated luminosity is
200~pb$^{-1}$. }
\label{fig-asy-gl}
\end{figure}

Fig.~\ref{fig-asy-gl} shows the expected measurable asymmetries for 
different sets of polarised gluon densities, i.e.~the ones shown
in Fig.~\ref{input}: GS-A, GS-C, and the instanton-gluon. 
For the latter the polarised
quark densities were taken from~\cite{GS}.
The assumed luminosity is 
200~pb$^{-1}$. It can be seen here that the measurable asymmetry is
very sensitive to the gluon input. The negative instanton-$\Delta G$  leads to
a positive asymmetry and can be clearly distinguished from the other two sets.
GS-A and GS-C can be discriminated in the higher $x$-range, which is where 
they are maximally different.

\section{Extraction of {\boldmath $\Delta G$} }

\vspace{1mm}
\noindent
In this section we will quantify the sensitivity to the shape of  $\Delta G/G$
and discuss systematic uncertainties.
In a real measurement one could obtain $\Delta G/G$ from the measured
asymmetry by an unfolding method, where the background would  be subtracted
statistically and correlations between bins are fully taken into account.
Such a method was used by H1 to extract the unpolarised
gluon density~\cite{H1}.
If correlations between bins are small one can use a simpler method
 performing a bin-by-bin correction. For our
study  we consider the latter method to be sufficient.

We simulate  500~pb$^{-1}$ of  di-jet events, as described 
in Sect.~3 with 
GS-A as input gluon density. This  would in a real measurement
correspond to the Monte Carlo generation of events and will therefore 
be called
 'MC-set' here.
Assuming that for each $x$-bin 
$\Delta G/G$ and 
$A_{corr}$ are related to each other  by a simple factor $F_i$:
$$ \left(\frac{\Delta G}{G}\right)_i = F_i \cdot A_{corr,\; i},$$
where $i$ indicates the $x$-bin,
 we compute
these factors using the  MC-set. 

\begin{figure}[t]
\epsfxsize=9cm
\hfil \epsffile{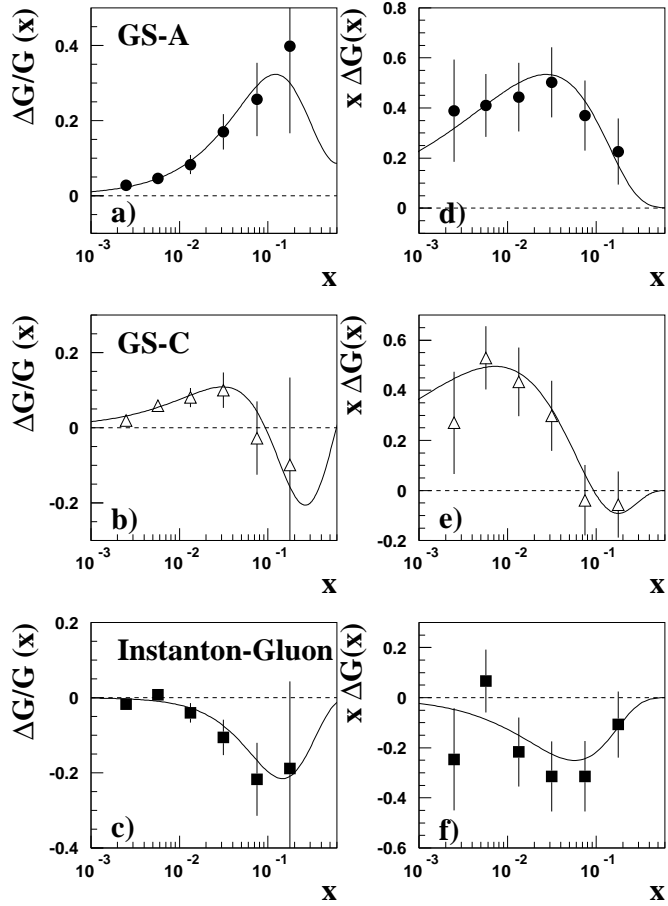} \hfil

\vspace{-1cm}
 \hspace*{12cm}
\caption{For a luminosity of 200~pb$^{-1}$ the sensitivity to extract
$\Delta G/G$ (a-c) and $x \Delta G$ (d-f) 
is shown for different polarised gluon densities
(see text).}
\label{fig-g-200}
\end{figure}

These factors $F_i$ were then multiplied with
 the asymmetries $A_{corr}$ that
correspond to the three measured asymmetries in Fig.~\ref{fig-asy-gl}. 
The three  sets of events used here represent the possible
measurements and are called  'data sets'.
The result is shown in Figs.~\ref{fig-g-200}a - \ref{fig-g-200}c.
Within the statistical accuracy the input
(solid lines)  
$\Delta G/G$ is found back for all cases
 and the statistics is sufficient to  
discriminate between them. The six  $x$-points
allow in particular a measurement of the
 shape of the polarised gluon distribution.
 (The statistical fluctuations for GS-A are smaller than can be expected 
because the data set of 200~pb$^{-1}$ which was used to produce  the
asymmetry was also included in the MC-set used to determine
the correction factors $F_i$.) 
The errors shown reflect the statistics of the data sets (200~pb$^{-1}$).
 The 
statistical uncertainty of the $F_i$ is not included here, since
in a real measurement it would be computed with very high statistical 
precision.
However, the limited statistics here is reflected in the 
fluctuations in Fig.~\ref{fig-g-200}.
Figures~\ref{fig-g-200}d - \ref{fig-g-200}f shows the same result presented
for the theoretically more interesting quantity $x \Delta G$. The
error bars are scaled errors of the left column, i.\ e.\ no uncertainty
was assigned to $G$ at this stage. \\

\begin{figure}[t]
\epsfxsize=9cm
\hfil \epsffile{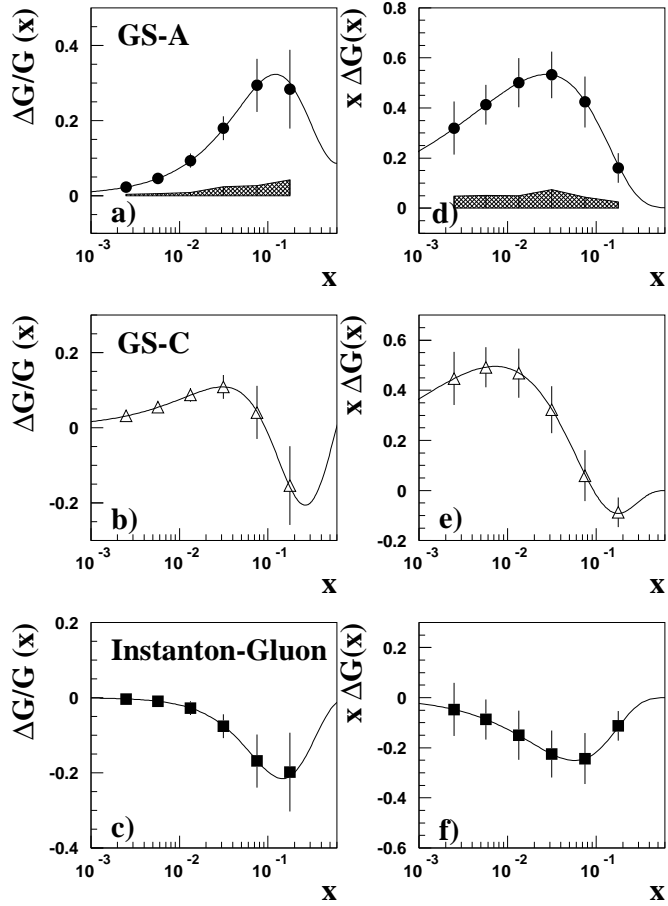} \hfil

\vspace{-1cm}
 \hspace*{12cm}
\caption{Sensitivity to $\Delta G/G$ (a-c) and $x\Delta G$ (d-f)
 for a luminosity
of 500~pb$^{-1}$.}
\label{fig-g-500}
\end{figure}

After we have shown that the measurable asymmetries are sensitive
to the input of $\Delta G$ and that we are able to extract 
the polarised gluon densities in several $x$-bins, 
the sensitivity to the shape of 
$\Delta G/G$ and $x \Delta G$ for an integrated
luminosity of 500~pb$^{-1}$
 is shown in Fig.~\ref{fig-g-500}. The statistical errors
for the  $x$ points are shown on the curves 
for $\Delta G/G$ (a-c) and $x \Delta G$ (d-f).
Again we notice the good separation between the different distributions. 
A study was performed on the systematic errors
for the  GS-A polarised gluon distribution.
The error sources considered were:  an
uncertainty of 2\% of the calibration of the hadronic energy scale,
an error on the total unpolarised di-jet cross section 
$\sigma_{di\mbox{-}jet}$ of 2\% and an error
on the unpolarised gluon density $G
(x)$ of 
5\%~\cite{Klein}. We assume that the unpolarised quantities will be 
measured before with high statistical precision with the HERA high-luminosity
upgrade.
The uncertainty on
the ratio of  the polarised and unpolarised quark densities, $\Delta q/q$, 
was considered to be 10\%, based on present fixed target measurements
of $g_1(x)$,  and the error on the polarisation measurement
was taken to be 5\% for each beam. These contributions
were added in quadrature and the result is displayed as a shaded band in the
Fig.~\ref{fig-g-500}a and~\ref{fig-g-500}d.
The largest contribution is due to the uncertainty on the beam polarisations 
and,
for the two highest $x$-bins, due to the QCDC contribution. 
Other studies, such as the influence of the choice of the fragmentation model
on the result, have also been performed, but no significant change of the
results could be observed. 
In summary we see that for all $x$ bins the statistical uncertainty is dominating.

In Table~\ref{errors} the expected statistical errors,
 corresponding
to a luminosity of 500~pb$^{-1}$, are detailed for the measurable $x$-bins
and for the first moment $\int \Delta G dx$ in the  range of
$0.0015 < x < 0.32$. The result we obtain is: 
\begin{equation}
\delta \left (\int_{0.0015}^{0.32} \Delta G dx\right ) = 0.21.
\end{equation}
The table also shows an interesting combination of this result with
a proposed measurement of $\Delta G/G$ from prompt photon production of
jets in polarised $pp$-scattering, called HERA-$\vec{N}$~\cite{HERAN}.
 The idea is to
scatter the polarised protons from the  HERA collider 
on a polarised fixed proton target. The luminosity assumed for
this experiment is 240~pb$^{-1}$ corresponding to about 3 years
of running in parallel with the collider experiments. 
Such an experiment would cover an almost completely complimentary $x$-range.
Combining the two $x$-ranges leads to the result, which is shown
 in the last row of the table.
The individual results entering into the combinations are given on
top. The covered $x$-range is increased to $0.0015 < x < 0.5$ and
the result on the error of the first moment of $\Delta G$ is:
\begin{equation}
\delta \left (\int_{0.0015}^{0.5} \Delta G dx \right ) = 0.20.
\end{equation}

\begin{figure}[t]

\epsfxsize=12cm
\rotate[r]{ \epsffile{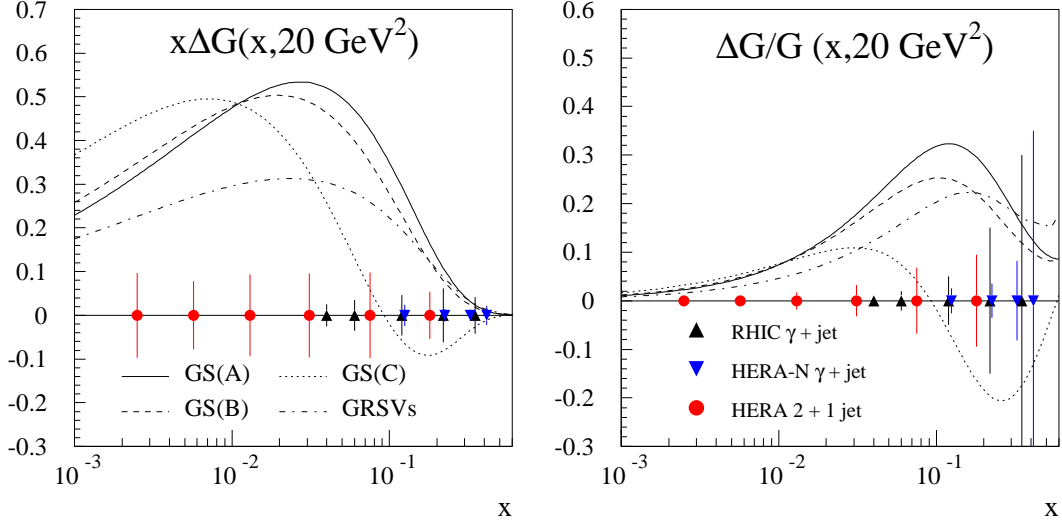}}

\vspace{-3.0cm}
 \hspace*{12cm}
\caption{The $x_g$-range accessible to different experiments for
measuring $\Delta G$. Different parametrizations of the polarised 
gluon density are shown as well, to demonstrate the separation
power of the measurements. } 
\label{allex}
\end{figure}

A comparison of the accessible $x$-ranges for this and other proposed
experiments is also shown in Fig.~\ref{allex}. Displayed is  the
expected di-jet result from polarised $ep$ scattering at HERA
(HERA 2+1 jet), the 
HERA-$\vec{N}$ $\gamma$+jet measurement, and the expected accuracy
from a measurement of $\gamma$+jet in polarised $pp$ collisions
with the STAR detector at  RHIC~\cite{RHIC} for $Q^2=20~{\rm GeV}^2$.
The di-jet measurement at a polarised 
HERA clearly extends into a region of $x$ which is not accessible to
any other experiment! Also shown  are four different
parametrizations for the polarised gluon densities. In addition to the
previously used 'gluons sets A and C' of Gehrmann and Stirling
also the 'gluon set B' of the same authors and the 'standard scenario'
of Gl\"uck, Reya, Stratmann and Vogelsang (GRSVs)~\cite{GRSV} (all  LO) are shown.
All these parametrizations are in agreement with present data. 

\begin{table}[H]
\begin{center}
\begin{tabular}{||c|c|c||}
\hline\hline
$x$ &$\delta(\Delta G(x)/G(x))$ & $\delta(\int \Delta G dx)$  \\
\hline \hline
0.0015 - 0.0036 & 0.007  & 0.082 \\
0.0036 - 0.009  & 0.009 & 0.072   \\
0.009  - 0.022  & 0.002 & 0.092   \\
0.022  - 0.054  & 0.003 & 0.098 \\
0.054  - 0.13   & 0.068 & 0.099\\
0.13   - 0.32   & 0.095 & 0.058 \\
\hline
$\Sigma$   0.0015 - 0.32   & 0.019  & 0.21 \\
\hline\hline
$\Sigma$ 0.0015 - 0.1 (di-jets) & 0.006  & 0.20 \\
$\Sigma$ 0.1 - 0.5 (HERA-$\vec{N}$)  & 0.017  & 0.021 \\
\hline
$ \Sigma$  0.0015 - 0.5  (di-jets + HERA-$\vec{N}$)& 0.018  & 0.20 \\  
\hline \hline
\end{tabular}
\end{center}
\caption{Expected uncertainties on $\Delta G/G$ and the first moment of 
$\Delta G(x)$ for the measured $x$-bins.
The upper seven  rows show  errors for each
 $x$-bin and the sum over the full measured range of the di-jet analysis.
In the lower three rows the di-jet measurement has been combined with
a possible measurement of $\Delta G/G$ from prompt photon + jet
production
at HERA-$\vec{N}$.}
\label{errors}
\end{table}

The first
moments of all the Gehrmann-Stirling sets are very similar (2.6, 2.6, and
2.5 at $Q^2=20~{\rm GeV}^2$ for the sets A, B, C, respectively, and 1.8 for
GRSVs).
The shape of $\Delta G(x)$, however, can be very different, which shows
the importance of this kind of measurement with respect to e.g. extractions
of the first moment of $\Delta G$ in a NLO-QCD analysis of the polarised
structure function $g_1$. However, since these two approaches
give rather complementary information it could be advantageous to combine
the di-jet analysis and the NLO fits to $g_1$ into a common fit. A case
study of such an analysis for a polarised HERA has been 
performed~\cite{comfit}.\\
Another point to stress here is that for all four
 parametrizations the part of the first moment $ \int_{0.0015}^{0.32} \Delta G dx$ which
can be measured with the di-jet events is  60\% for GS-C and about 75\%
for GS-A, GS-B, GRSVs of the total first moment $\int_{0}^{1} \Delta G dx$.
As an example, assuming GS-A, in this experiment we would measure
$\int_{0.0015}^{0.32}\Delta G dx = 2.0 \pm 0.21 ({\rm stat.})$, hence a 
10\% uncertainty for the first moment in the measured range.

\begin{figure}[t]
\epsfxsize=9cm
\hfil \epsffile{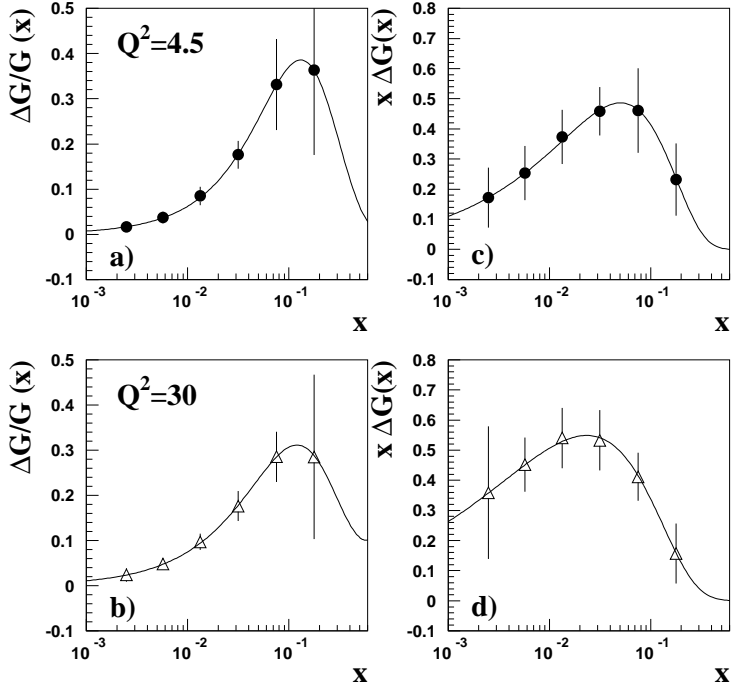} \hfil
\vspace{-2.5cm}
 \hspace*{12cm}
\caption{Sensitivity to $\Delta G/G$ (a,b) and $x\Delta G$ (c,d) 
when splitting the data
into two bins of $Q^2$. The low-$Q^2$ cut was reduced to $Q^2>2~{\rm GeV}^2$.
The assumed  luminosity is 500~pb$^{-1}$.}
\label{fig-g-q2}
\end{figure}
 
To show that  more information  could  be
extracted, Fig.~\ref{fig-g-q2} shows
the expected statistical uncertainty on $\Delta G/G$ (a,b) and $x \Delta G$
(c,d) for two bins of $Q^2$. A luminosity of 500~pb$^{-1}$ and the GS-A 
polarised gluon distribution are assumed. The data were divided as 
given in 
 Table~\ref{tab-asys}. It shows  that such analysis can provide direct 
information on the interesting question 
on the $Q^2$ dependence of $\Delta G$.

\section{Conclusions}

\vspace{1mm}
\noindent
We have shown in this study that an analysis of the 
di-jet rate at HERA allows 
a measurement of $\Delta G/G(x)$ 
in an $x$-range from $0.002 < x < 0.2$, a region where 
large differences are observed between present models for 
the polarised gluon distribution. 
This  $x$ range is largely uncovered 
 by any other proposed experiment. The precision of the 
measurement, both statistical and 
systematical, is large enough such that shape of $\Delta G/G$
could be measured  and discrimination 
between different polarised gluon distributions would be possible.
The first moment of $\Delta G$ can be determined with a precision of
about 10\%, in the range $0.0015 < x < 0.32.$
The results are complimentary to extractions of the first moment of $\Delta G$
from structure function measurements, 
and measurements at COMPASS~\cite{COMPASS}, RHIC or HERA-$\vec{N}$.
The
proposed measurement is vital for  our understanding  
of the spin structure  of the 
nucleon.

\section*{Acknowledgments}

\vspace{1mm}
\noindent
 We thank 
T. Gehrmann, F. Kunne and E. Mirkes  for helpful discussions.

\noindent
``These Proceedings'' refers to the Proceedings of the Workshop on Physics
at HERA with Polarized Protons and Electrons.

\end{document}